\documentclass[useAMS,usenatbib]{mn2e}
\usepackage{graphicx,amssymb, txfonts}

\def\pb{Pa$\beta$}
\def\br{Br$\gamma$}

\def\feii{[Fe\,{\sc ii}]}
\def\pii{[P\,{\sc ii}]}
\def\oiii{[O\,{\sc iii}]}

\def\h2{H$_2$}
\def\p1{Paper~I}

\def\kms {$\rm km\,s^{-1}$}

\title[Feeding vs. Feedback in Mrk~79]{Feeding Versus Feedback in AGNs from Near-Infrared IFU Observations: The Case of Mrk~79}

\author[Riffel, Storchi-Bergmann \& Winge]{Rogemar. A. Riffel$^{1}$\thanks{E-mail:
rogemar@ufsm.br}, Thaisa Storchi-Bergmann$^{2}$ and Claudia Winge$^{3}$\\
$^{1}$ Universidade Federal de Santa Maria, Departamento de F\'\i sica, Centro de Ci\^encias Naturais e Exatas, 
97105-900, Santa Maria, RS, Brazil\\
$^{2}$ Universidade Federal do Rio Grande do Sul, Instituto de F\'\i sica, CP 15051, Porto Alegre 91501-970, RS, Brazil\\
$^{3}$ Gemini Observatory, c/o Aura, Inc., Casilla 603, La Serena, Chile
}
\begin{document}


\pagerange{\pageref{firstpage}--\pageref{lastpage}} \pubyear{2011}

\maketitle

\label{firstpage}

\begin{abstract}

We have mapped the gaseous kinematics and the emission-line flux distributions and ratios from the inner $\approx$\,680\,pc radius of the Seyfert 1 galaxy Mrk\,79, using two-dimensional (2D) near-IR $J-$ and $K_l-$band spectra obtained with the Gemini instrument NIFS at a spatial resolution of $\approx$100~pc and velocity resolution of $\approx$\,40\,\kms. 
The molecular hydrogen H$_2$ flux distribution presents two spiral arms extending by $\approx$\,700\,pc, one to the north and another to the south of the nucleus, with an excitation indicating heating by X-rays from the central source. The low velocity dispersion ($\sigma\approx50$\kms) and rotation pattern supports a location of the H$_2$ gas in the disk of the galaxy. Blueshifts observed along the spiral arm in the far side of the galaxy and redshifts in the spiral arm in the near side, 
suggest that the spiral arms are feeding channels of H$_2$ to the inner 200\,pc. From channel maps
along the  H$_2\,\lambda2.1218\,\mu$m emission-line profile we estimate a mass inflow rate of $\dot{M}_{H_2}\approx4\times10^{-3}\, {\rm M_\odot\,yr^{-1}}$, which is one order of magnitude smaller than the mass accretion rate necessary to power the AGN of Mrk\,79.
The emission from the ionized gas (traced by \pb\ and \feii$\,\lambda1.2570\mu$m emission lines) is correlated with the radio jet and with the narrow-band  \oiii\ flux distribution. 
Its kinematics shows both rotation and outflows to the north and south of the nucleus.
The ionized gas mass outflow rate through a cross section with radius $\approx$\,320\,pc located at a distance of $\approx$455\,pc from the nucleus is $\dot{M}_{\rm out}\approx3.5~{\rm M_\odot\, yr^{-1}}$,  which is much larger than the AGN mass accretion rate, indicating that most of the outflowing gas originates in the interstellar medium surrounding the galaxy nucleus, which is pushed away by a nuclear jet. 

\end{abstract}


\section{Introduction}

This work is part of a large project, in which our group (Active Galactic Nuclei Integral Field
Spectroscopy - AGNIFS) has been observing the inner kiloparsec of nearby active galaxies using
optical and near-infrared (hereafter near-IR) high-spatial resolution (a few to tens of pc)  integral field spectroscopy. We have obtained the gaseous flux distributions, kinematics and excitation, with the main goal of mapping gas inflows and outflows and constraining the corresponding mass flow rates and power. Whenever the signal-to-noise ratio in the continuum and absorption lines is high enough we also map the stellar kinematics and ages. In the near-IR, our main results so far can be summarized as follows. We have found that the  molecular (\h2) and ionized gases present distinct flux distributions and kinematics, with the former restricted to the plane of the galaxy and presenting, in some cases, streaming motions towards the nucleus, while the latter presents emission from outflowing material at high latitudes above the plane and usually is associated to the radio emission \citep{riffel06,n4051,n7582,mrk1066a,mrk1066c,sb09,sb10}. We have concluded that the molecular gas is  a good tracer of the nuclear feeding and the ionized gas of its feedback. Nevertheless, the sample observed so far comprises only half a dozen objects and more integral field observations in the near-IR  are required in order to have a more complete census of these processes in AGNs  and  to relate the mass flow rates to the AGN power.

In this work, we present the gaseous flux distribution and kinematics of the inner 680 pc of the active galaxy Mrk\,79 obtained from observations using the Gemini-North's Near-Infrared Integral Field Spectrograph \citep[NIFS][]{mcgregor03} in the J and K-bands.  This object was selected for this study because it presents extended radio and [O\,{\sc iii}] emission \citep[e.g.][]{ulvestad84,nagar99,schmitt03} allowing us to explore the relation between the radio jet and the Narrow Line Region (NLR) kinematics, as well as its effect in the excitation of the near-IR lines.

Mrk\,79 is a SBb galaxy \citep{devaucouleurs91} located at a distance of 93.8~Mpc \citep[e.g.][]{kraemer11}, for which 1 arcsec corresponds to 455\,pc at the galaxy. Its nucleus is classified as Seyfert 1 with a central black hole with a mass of $5.2\pm1.44\times10^7\,{\rm M_\odot}$ \citep{peterson04}. It presents extended radio-continuum emission along position angle (PA) 11$^\circ$  \citep{schmitt01,nagar99,ulvestad84}, which can be described as an asymmetric triple radio structure with the northern source being located at a distance of 800\,pc from the nucleus and the southern source at 460\,pc from it \citep{schmitt01}. In the optical, Mrk\,79 presents extended [O\,{\sc iii}] line emission as observed in ground-based \citep[e.g.][]{haniff88} and Hubble Space Telescope (HST) images \citep{schmitt03}. The [O\,{\sc iii}]$\lambda5007$ emission extends to about 4.6 arcsec from the nucleus in the north-south direction, being co-spatial with the radio emission and presenting several blobs of enhanced emission \citep{schmitt03}. An outflowing gas component is suggested by the detection of two components in the [O\,{\sc iii}]$\lambda5007$ mission-line profile with velocities of $+100$\,km\,s$^{-1}$ and $-50$\,km\,s$^{-1}$ relative to the systemic velocity of the galaxy. This component seems to be associated with the northern radio structure \citep{whittle88}. The modeling of the optical and infrared spectral energy distribution (SED) of Mrk\,79, including the emission by dust in a toroidal-like structure heated by a central AGN suggests that the dusty torus has a mass of $7.30\times10^2 M_\odot$  \citep{fritz06}.

\begin{figure*}
 \centering
 \includegraphics[scale=0.8]{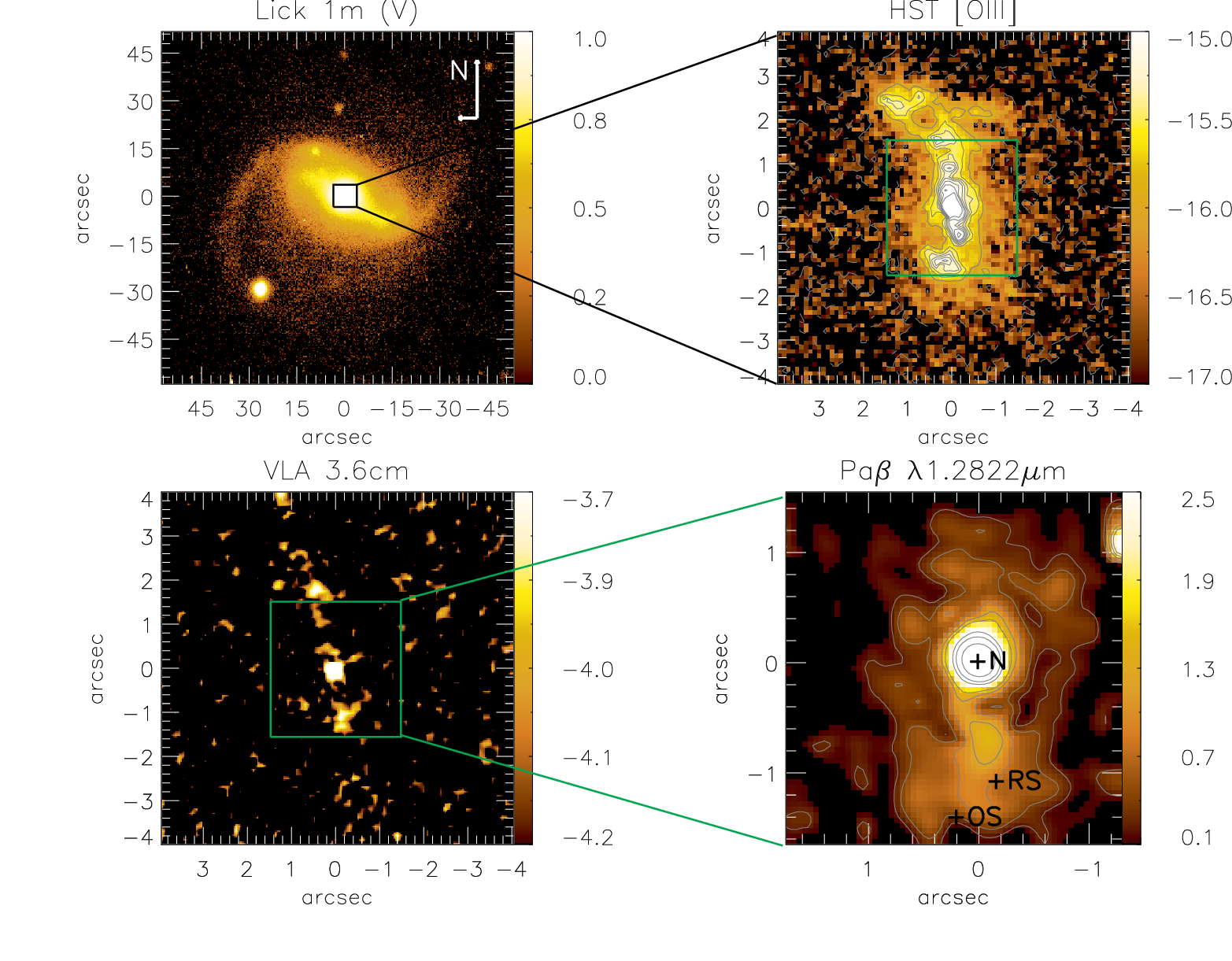} 
  \caption{Top-left: V-band optical image of Mrk\,79 obtained with the with the 1-m telescope of the Lick Observatory \citep{hunt99}.  Top-right: \oiii$\,\lambda5007\,\AA$ image obtained with the  HST \citep{schmitt03}. Bottom-left: 3.6\,cm radio continuum image obtained with the VLA by \citet{schmitt01}. The green box overlaid to this panel represents the NIFS field of view. Bottom-right: \pb\ flux map obtained from our NIFS datacube in units of 10$^{\rm -17}{\rm \,erg\,s^{-1}\,cm^{-2}}$. The positions labeled as N, RS and OS mark the location of the nucleus, the southern radio spot and the southern [O\,{\sc iii}] southern blob, respectively.} 
 \label{large}  
 \end{figure*}

\begin{figure*}
 \centering
 \includegraphics[scale=0.8]{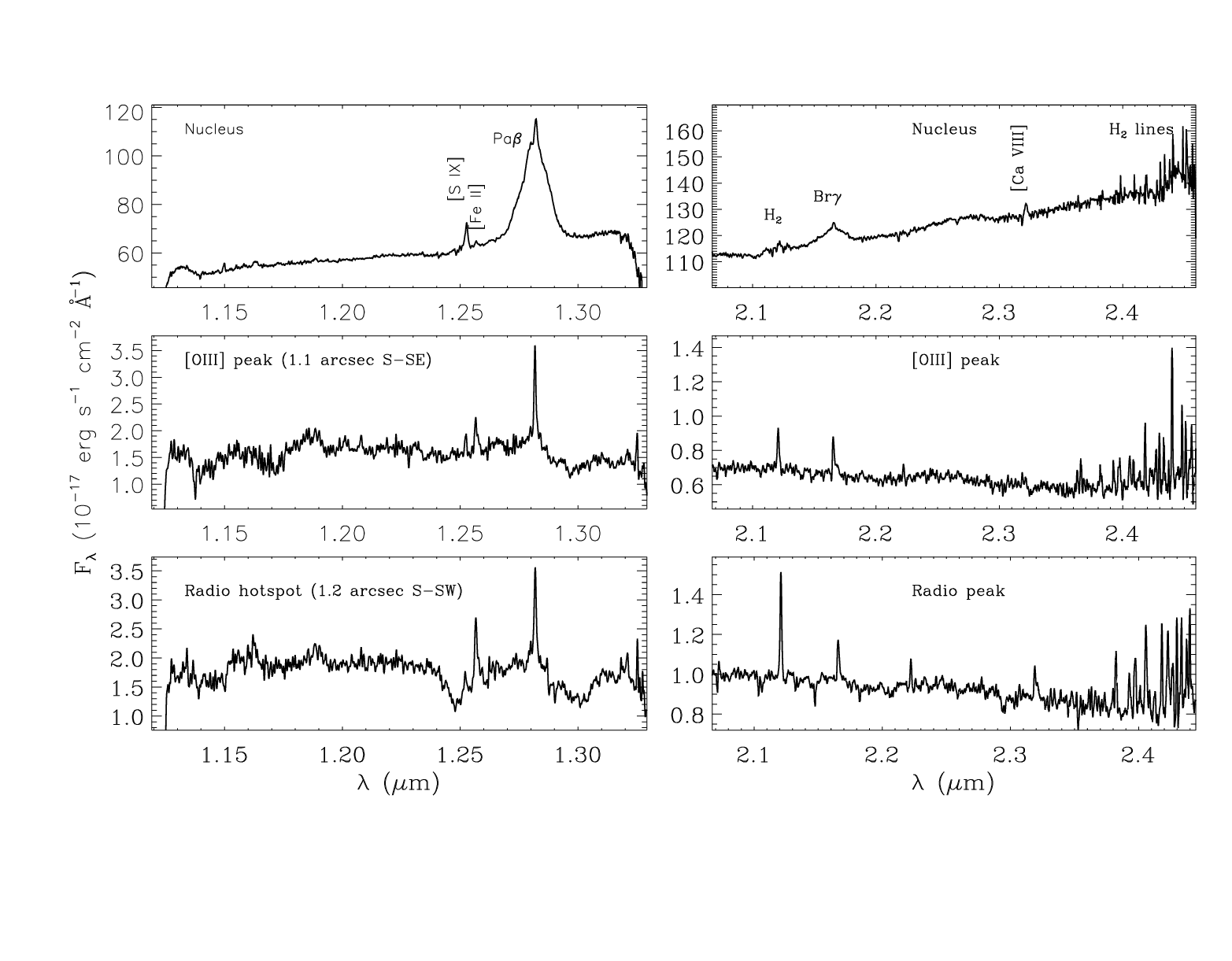} 
  \caption{Sample of spectra for Mrk\,79 extracted within square apertures of 0\farcs25$\times$0\farcs25, with the stronger emission-lines identified. Top panels show the nuclear J (left) and K (right) spectrum, middle panels show a spectrum for a location at 1\farcs2 south-southeast of the nucleus, co-spatial with an [O\,{\sc iii}] emission knot and the bottom panels present a spectrum extracted at the position of the radio structure at 1\farcs2 south-southwest of the nucleus. These positions are identified in figure \ref{large} as "N", "OS" and "RS", respectively.} 
 \label{espec}  
 \end{figure*}

This paper is organized as follows. In Sec.~\ref{obs} we describe the observations and data reduction procedures. The results are presented in Sec.~\ref{results} and discussed in Sec.~\ref{discussion}. We present our conclusions in Sec.~\ref{conclusions}.

\section{Observations and Data Reduction}\label{obs}

The J and Kl-band observations of Mrk 79 were obtained using the NIFS instrument \citep{mcgregor03} operating with the Gemini North Adaptive Optics system ALTAIR in September 2010 under the programme GN-2010B-Q-42, following the standard  Object-Sky-Sky-Object dither sequence, with off-source sky positions since the target is extended, and individual exposure times of 520\,s for the J band and 550\,s for the K$_l$ band. Six on-source individual exposures were obtained for each band, totalizing 3120\,s and 3300\,s for the J and K$_l$-band, respectively.

 The J-band observations covered the spectral region from 1.14\,$\mu$m to 1.36\,$\mu$m, centered at 1.25\,$\mu$m with a spectral resolution of $\approx1.8\,\AA$, as obtained from the measurement of the full width at half maximum (FWHM) of the ArXe calibration lamp lines. The K$_{\rm l}$-band observations were centered at 2.3\,$\mu$m, covering the spectral range from  2.10$\,\mu$m to 2.53$\,\mu$m with a spectral resolution of FWHM$\approx3.5\,\AA$. In velocity space, the resolution of the observations is $\approx$\,45\kms\ for the K$_{\rm l}$-band and 35 \kms\ for the J-band.

The data reduction was accomplished using tasks contained in the {\sc nifs} package which is part of {\sc gemini iraf} package, as well as generic {\sc iraf} tasks. The reduction procedure included trimming of the images, flat-fielding, sky subtraction, wavelength and s-distortion calibrations. We have also removed the telluric bands and flux calibrated the frames by interpolating a black body function to the spectrum of the telluric standard star. The six individual datacubes of each band were median combined to a single datacube using the {\it gemcombine} task of the {\sc gemini iraf} package, with a {\it sigclip} algorithm to eliminate the remaining cosmic rays and bad pixels. The final IFU data cube in each band contains $\sim4000$ spectra, each spectrum corresponding to an angular coverage of 0$\farcs$05$\times$0$\farcs$05, which translates into $\sim$23$\times$23\,pc$^2$ at the galaxy and covering the inner 3\arcsec$\times$3\arcsec ($\sim$1.35$\times$1.35\,kpc$^2$).

In order to increase the signal to noise ratio and allow the fitting of the emission line profiles, we have replaced each spatial pixel by the median of its value and that of  its first 8 neighbors. After this procedure, the angular resolution, obtained  from the FWHM of the spatial profile of the broad components of \br\ and \pb\ emission lines fluxes, is  0\farcs25$\pm$0\farcs05 for both bands, corresponding to $\sim$100\,pc at the galaxy.

\section{Results}\label{results}

In order to illustrate the spatial coverage of our observations, we show in the top-left panel of Figure\,\ref{large} a V-band optical image of Mrk\,79 obtained with the with the 1-m telescope of the Lick Observatory \citep{hunt99}. In the top-right panel we present an \oiii$\,\lambda5007\,\AA$ narrow-band image obtained with the  HST Wide Field Planetary Camera 2 (WFPC2) through the filter FR533N \citep{schmitt03}.  This image shows extended emission up to 4\arcsec from the nucleus of Mrk\,79, being more elongated along position angle (PA) 15$^\circ$ and presenting several blobs of emission. \citet{schmitt03} suggest that the [O\,{\sc iii}] emission is related to the radio jet. The 3.6\,cm radio continuum image obtained with the Very Large Array by \citet{schmitt01} is shown in the bottom-left panel of this figure, and can be described as an asymmetric triple radio structure along PA$=$11$^\circ$. The green box overploted on this panel represents the NIFS field of view and shows that the north-eastern radio structure is beyond its borders. The \pb\ flux map obtained from our NIFS datacube is shown in the bottom-right panel in units of 10$^{\rm -17}{\rm \,erg\,s^{-1}\,cm^{-2}}$.

 All spectra in the data cube were first corrected for redshift. In Figure\,\ref{espec} we present the J and K$_l$ spectra extracted within an 0\farcs25$\times$0\farcs25 aperture centered on  the the nucleus in the top panels.  The middle panels show the spectra extracted at 1\farcs1 south-southeast of the  nucleus at the location where it is observed a knot with enhanced emission in the [O\,{\sc iii}] image in the top-right panel of Figure\,\ref{large}, while the bottom panels show the spectra for a location co-spatial with the southern radio hotspot seen in bottom-left panel of Figure\,\ref{large} at 1\farcs1 south-southwest of the nucleus. Some of the strongest emission lines are identified in the top panels.

We have detected about 4 dozen of emission lines at the J and K$_l$-bands as listed in Table\,\ref{fluxes_j}, which presents the corresponding fluxes of the lines for the three positions above. The observed emission lines include molecular (H$_2$) lines as well as lines of ionized gas from  a range of  ionization levels, e.g. from \feii\ up to [S\,{\sc ix}]. The line fluxes were measured by fitting a Gaussian to each emission-line profile using the {\sc iraf} {\sc splot} task. The uncertainties quoted in the table are the standard deviations from the average of several measurements (typically 6) and do not include uncertainties in the flux calibration of the spectra. Flux values followed by ``?" indicate that the line is only marginally detected, with a large uncertainty,  of the order of the line flux. The presence of the line is supported by the fact that it has the expected central wavelength and the fact that the width is similar to that of the other emission lines.

\subsection{Emission-Line Flux Distributions}\label{flux_distributions}

\begin{table*}
\centering
\caption{J-band emission-line fluxes for the nucleus, the position of an [O\,{\sc iii}] blob (at 1\farcs1 south-southeast of the nucleus) and  the position of the radio hotspot (at 1\farcs2 south-southwest of the nucleus)  integrated within 0\farcs25$\times$0\farcs25 apertures. The location where each spectrum was extracted is identified in the bottom-left panel of Fig.\,\ref{large} as N (for the nucleus), OS ([O\,{\sc iii}] structure) and RS (radio structure). The fluxes are in $10^{-17}$erg s$^{-1}$cm$^{-2}$ units.  A "?" by the side of the flux value indicates that the line is detected but the uncertainty in the measurement is of the order of the flux of the line.}
\vspace{0.3cm}
\begin{tabular}{l l c c c }
\hline
\hline
$\lambda_{vac} {\rm(\mu m)}$   & ID                  &Nucleus           &[O\,{\sc iii}] peak & Radio peak	\\
\hline
1.12708  & [Fe\,{\sc ii}]\,$b^2F_{5/2}-b^4F_{5/2}$   &      --  	& --		&  2.45$\pm$1.18 \\
1.13488  & [Fe\,{\sc ii}]\,$b^2D_{3/2}-a^2F_{7/2}$   &      --  	& 1.54? &  3.17$\pm$0.84 \\
1.14276  & [Fe\,{\sc ii}]\,$b^4D_{5/2}-b^4F_{9/2}$   &      --  	& --		&  0.61? \\
1.14713  & [P {\sc ii}]\,$^1D_3-^3P_1$               &  32.39$\pm$1.97  & --		&  1.87? \\
1.15841  & [Fe\,{\sc ii}]\,$b^2F_{5/2}-a^2D_{3/2}$   &  7.13?             &  1.95?  \\
1.16296  & He\,{\sc ii}\,$7-5$                       &  43.71$\pm$7.75  & --		&  5.23$\pm$1.51  \\
1.18363  & [Fe\,{\sc ii}]\,$b^4D_{5/2}-b^4F_{5/2}$   &  5.29?  & 0.76? &  -- \\
1.18552  & He\,{\sc ii}\,$29-7$                      &  --		& 1.28? &  -- \\
1.18693  & [Fe\,{\sc ii}]\,$b^4D_{3/2}-b^4F_{5/2}$   &      --  	& 1.44? & 	 --	  \\
1.18861  & [P {\sc ii}]\,$^1D_2-^3P_2$               &  17.30$\pm$5.85  & 3.00$\pm$0.28 &  4.57$\pm$0.77  \\
1.20370  & He\,{\sc ii}\,$26-7$                      &  --		& 0.77? &  1.18? \\
1.20545  & [Fe\,{\sc ii}]\,$a^4D_{3/2}-a^6F_{7/2}$   &   5.73?  & -- & 	 --	  \\
1.22263  & [Fe\,{\sc ii}]\,$a^4D_{1/2}-a^6D_{5/2}$   &  8.78?  & 2.02? &  1.19? \\
1.24414  & [Fe\,{\sc ii}]\,$c^2G_{9/2}-a^4G_{11/2}$  &        --	& --            &  0.97? \\
1.25235  & [S\,{\sc ix}]\,$^3P_1-^3P_2$              & 161.97$\pm$7.91  & 3.99$\pm$0.25 &  2.55$\pm$0.71 \\
1.25702  & [Fe\,{\sc ii}]\,$a^4D_{7/2}-a^6D_{9/2}$   &  26.79$\pm$7.33  & 8.12$\pm$1.99 &  13.20$\pm$0.29 \\
1.27069  & [Fe\,{\sc ii}]\,$a^4D_{1/2}-a^6D_{1/2}$   &  --		& 1.32? &   -- \\
1.27912  &[Fe\,{\sc ii}]\,$a^4D_{3/2}-a^6D_{3/2}$/HeI? &   --		& 1.39? &  0.71? \\   
1.28216  &  H\,{\sc i}\,Pa$\beta$ (broad)            &5666.25$\pm$164.06& --		& 	  --	   \\
1.28216  &  H\,{\sc i}\,Pa$\beta$ (narrow)           & 154.28$\pm$15.41 &16.34$\pm$0.61 &  16.56$\pm$1.04   \\
1.28495  &  He\,{\sc i}\,$^3S_1-^3Po_{1}$            &  --		& 1.40? &  0.87? \\
1.29812  & [Fe\,{\sc ii}]\,$a^4D_{3/2}-a^6D_{1/2}$   & 11.85$\pm$3.30  & --		&  0.39?  \\
1.30529  & [Fe\,{\sc ii}]\,$c^2G_{9/2}-a^4G_{9/2}$   &  --		& 1.46? &  0.59? \\
1.31958  & [Fe\,{\sc ii}]\,$b^2G_{7/2}-b^4F_{9/2}$   &  --		& -- &  1.36? \\
1.32092  & [Fe\,{\sc ii}]\,$a^4D_{7/2}-a^6D_{7/2}$   &  --		& 2.46? &  3.50? \\

2.12183  & H$_2$\, 1-0\,S(1) 	  		     &  89.89$\pm$23.19& 4.42$\pm$0.09 &  8.85$\pm$0.74 \\
2.15420  &  H$_2$\,1-0\,S(2)  	                     &  --		&   --  	&  0.80? \\
2.16612  & H\,{\sc i}\,Br$\gamma$(broad)             &1054.40$\pm$159.49&  --		& 	  --	  \\
2.16612  & H\,{\sc i}\,Br$\gamma$ (narrow)           &  48.96$\pm$9.36 & 4.08$\pm$0.72 &  3.21$\pm$0.27 \\
2.17661  & [Fe\,{\sc ii}]\,$b^2G_{7/2}-b^2H_{11/2}$  &  22.06?&  --		& 	  --	  \\
2.18911  & He\,{\sc ii}\,$10-7$                      &  --		& 1.53$\pm$0.08 & 	 --	  \\
2.22344  &  H$_2$\,1-0\,S(0)  	                     &  19.51$\pm$6.74 & 1.35$\pm$0.13 &  2.14$\pm$0.35 \\
2.24776  &  H$_2$\,2-1\,S(1)  	                     &  		& --		&  1.39$\pm$0.34 \\
2.26647  & [Fe\,{\sc ii}]\,$b^2G_{7/2}-b^2H_{9/2}$   &  --		& 0.59? & 	  --	  \\
2.32204  &[Ca\,{\sc viii}]\,$^2P^0_{3/2}-^2P^0_{1/2}$&  165.60$\pm$38.37& --		&  3.50$\pm$0.65 \\
2.36760  & [Fe\,{\sc ii}]\,$a^4G_{9/2}-a^4H_{9/2}$   &  --		& 2.43? & 	  --	  \\
2.39901  & [Fe\,{\sc ii}]\,$b^4D_{7/2}-a^2F_{7/2}$   &  --		& 3.90$\pm$0.21 &  4.79? \\
2.40847  &  H$_2$\,1-0\,Q(1)   	                     &  --		& 2.75$\pm$0.22 &  6.81$\pm$0.50 \\
2.41367  &  H$_2$\,1-0\,Q(2)   	                     &  --		& 1.01? & 	  --	  \\
2.42180  &  H$_2$\,1-0\,Q(3)   	                     &  --		& 4.28$\pm$0.06 & 5.20$\pm$0.21  \\
2.43697  &  H$_2$\,1-0\,Q(4)   	                     &  --		& 2.61? & 	  --	  \\
2.46075  & He\,{\sc ii}\,$18-9$                      &  --		& 8.20$\pm$1.41 & 	  --	  \\	
									        				   
\hline

\end{tabular}
\label{fluxes_j}
\end{table*}

\begin{figure*}
 \centering
 \includegraphics[scale=0.8]{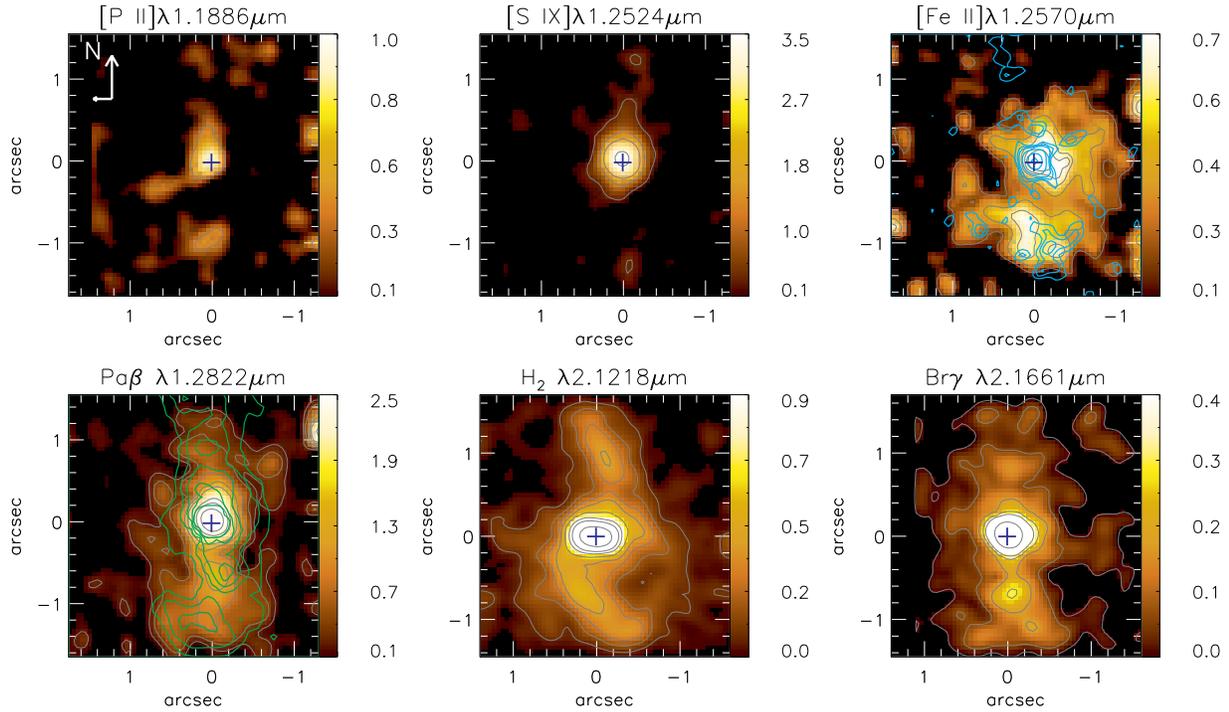} 
  \caption{Emission-line flux distributions for [P\,{\sc ii}]\,$\lambda$1.1886\,$\mu$m, [S\,{\sc ix}]\,$\lambda$1.2526\,$\mu$m, [Fe\,{\sc ii}]\,$\lambda$1.2570\,$\mu$m, Pa$\beta$, H$_2\,\lambda$2.1218$\mu$m and \br\ emission lines. The color bars show the flux scale in units of $10^{-17}$erg s$^{-1}$cm$^{-2}$. The central cross marks the position of the nucleus, the cyan contours overlaid to the [Fe\,{\sc ii}] map are from the 3.6~cm radio-continuum image of \citet{schmitt01} and the green contours overlaid to the \pb\ map are from the [O\,{\sc iii}] image of \citet{schmitt03}.} 
 \label{flux}  
 \end{figure*}

In order to construct maps for the flux distribution of the strongest emission lines, we used the {\sc profit} routine \citep{profit} to integrate the fluxes under the profiles of [P\,{\sc ii}]\,$\lambda$1.1886\,$\mu$m, [S\,{\sc ix}]\,$\lambda$1.2526\,$\mu$m, [Fe\,{\sc ii}]\,$\lambda$1.2570\,$\mu$m, Pa$\beta$, H$_2\,\lambda$2.1218$\mu$m and \br\ emission lines and subtract the underlying continuum. These particular lines have been chosen because they have the highest signal-to-noise (S/N) ratios among their species (coronal lines, forbidden and permitted  ionized gas lines and molecular lines). We were unable to construct maps for the flux distributions of He\,{\sc i}, He\,{\sc ii} and [Ca\,{\sc viii}] emission lines because they are detected only at a few locations of the observed field. The broad components for \pb\ and \br\ were fitted and subtracted before the construction of the  flux maps in these emission lines.

Figure~\ref{flux} shows the resulting flux maps for each emission line (identified in the top of each panel). The central cross marks the position of the nucleus defined as the peak of the near-IR continuum emission, the cyan contours overlaid to the [Fe\,{\sc ii}] map are from the radio continuum image from \citet{schmitt01}, shown in the top-right panel of Fig.~\ref{large}. The green contours overlaid to the \pb\ flux map are from the [O\,{\sc iii}] image of \citet{schmitt03}, shown in the bottom-left panel of Fig.~\ref{large}. The lower flux level shown in each panel corresponds to 3$\sigma$, where $\sigma$ is the noise in the adjacent continuum to the line.

The coronal line [S\,{\sc ix}]\,$\lambda$1.2526\,$\mu$m (top-middle panel of Fig.\,\ref{flux}) emission is marginally
resolved by our observations (with FWHM$\approx$0\farcs35) being more extended along the north-south direction with flux peak at the nucleus. A similar behavior is observed for the [P\,{\sc ii}] flux distribution (top-left panel), with an additional emission from a region located at 1\farcs1 south of the nucleus, almost co-spatial with the southern radio hotspot seen in Fig.~\ref{large}.
The [Fe\,{\sc ii}]  emission is extended up to 1\farcs4 from the nucleus, showing two structures at the highest flux levels: one at the nucleus and another co-spatial with the southern radio hotspot. Lower level emission is observed surrounding these structures.

 The \pb\ and \br\ flux maps show extended emission up to the borders of the NIFS field to the north and to the south, while in the east-west direction it extends only to $\approx$0\farcs5 from the nucleus. The H\,{\sc i} emission seems to be well correlated with the [O\,{\sc iii}] structures, as evidenced by the contours of the [O\,{\sc iii}] image overlaid on the \pb\ flux map. Finally, the \h2\ also presents extended emission more elongated in the north-south direction, but with a distinct flux distribution than those observed for the other emission lines. The \h2\ map clearly shows two spiral arms extending up to 1\farcs5 arcsec from the nucleus, which seem to originate from the tips of a nuclear bar-shaped
structure oriented approximately along east-west, observed at the highest flux levels.

\begin{figure*}
 \centering
 \includegraphics[scale=0.64]{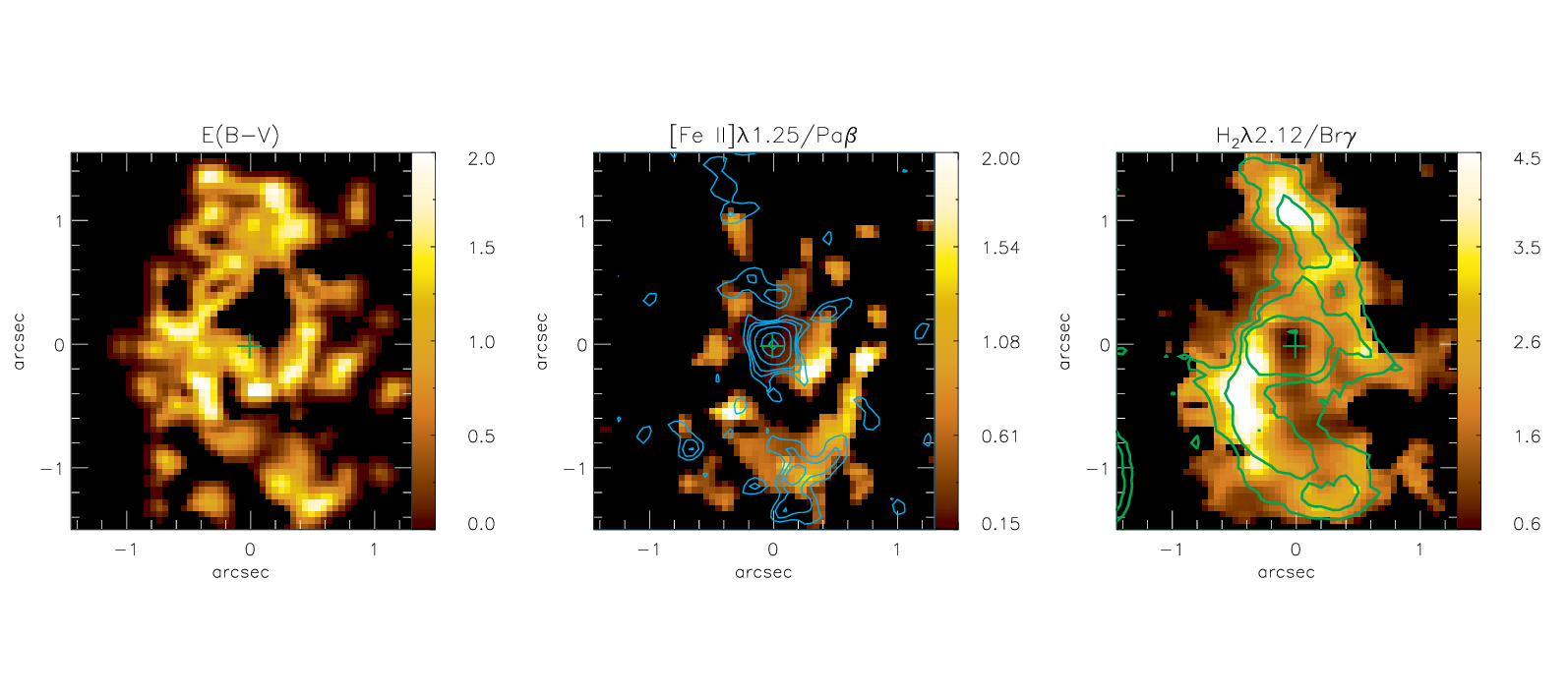} 
  \caption{Emission-line ratio maps. Left: E(B-V) obtained from Pa$\beta$/Br$\gamma$ line ratio. Middle: [Fe\,{\sc ii}]$\lambda$1.2570$\mu$m/\pb. Right: H$_2\lambda$2.1218$\mu$m/\br. The contours overlaid to the [Fe\,{\sc ii}]$\lambda$1.2570$\mu$m/\pb\ are for the radio image and to the  H$_2\lambda$2.1218$\mu$m/\br\ for the \h2\ flux map of Fig.~\ref{flux}.
 } 
 \label{ratio}  
 \end{figure*}

\subsection{Line-Ratio Maps}\label{ratio_maps}

Line ratios can be used to study the extinction and the excitation mechanisms of the near-IR lines in the NLR \citep[e.g.][]{dors12,mrk1157,mrk1066a,rogerio06,sb09,ardila05,ardila06}. 

The reddening can be estimated from the Pa$\beta$/Br$\gamma$ line ratio as

\begin{equation}
 E(B-V)=4.74\,{\rm log}\left(\frac{5.88}{F_{Pa\beta}/F_{Br\gamma}}\right),
\end{equation}
where $F_{Pa\beta}$ and $F_{Br\gamma}$ are the fluxes of $Pa\beta$ and $Br\gamma$ emission lines, respectively. We have used the reddening law of \citet{cardelli89} and adopted the intrinsic ratio $F_{Pa\beta}/F_{Br\gamma}=5.88$ corresponding to case B recombination \citep{osterbrock06}. The resulting $E(B-V)$ map is shown in the left panel of Fig.\,\ref{ratio}. This map shows a very complex structure with several knots of higher extinction, in which $ E(B-V)$ reaches values of up to 2.  

The excitation mechanism of the [Fe\,{\sc ii}] lines can be investigated using the [Fe\,{\sc ii}]$\lambda$1.2570$\,\mu$m/Pa$\beta$ line ratio, shown in the middle panel of Fig.~\ref{ratio}. 
Typical values for this ratio are [Fe\,{\sc ii}]$\lambda$1.2570$\,\mu$m/Pa$\beta\approx$0.6, observed over most of the field. 
The lowest values  are observed at the nucleus ($\approx$0.15), while the highest values of up to 2 are observed to the southwest at $\approx$0\farcs4 from the nucleus and in an arc-shaped structure at $\approx$1\arcsec. At the location of the southern ratio hotspot (cyan contours are from the radio image) the average value for this ratio is 1.2.

An useful line ratio to study the excitation mechanism of the near-IR H$_2$ lines is H$_2\lambda$2.1218$\,\mu$m/Br$\gamma$. We present a map for this ratio in the right panel of Fig.~\ref{ratio}. The highest values of up to 4.5 are observed along the two spiral arms seen in the H$_2$ flux map shown in the bottom-middle panel of Fig.~\ref{flux}, overlaid as green contours in the line-ratio map. The lowest values, around 0.6, are observed at the nucleus and at locations away from the spiral structures.

\subsection{Centroid Velocity and Velocity Dispersion Maps}

The {\sc profit} routine \citep{profit} was used to fit the emission line profiles by Gaussian curves in order to obtain the centroid velocity ($V$) and velocity dispersion ($\sigma$), which have been used to map of the gas kinematics in the inner region of Mrk\,79. We used the  [Fe\,{\sc ii}]$\lambda$1.2570$\mu$m, \pb\ and the H$_2\lambda$2.1218$\mu$m emission lines to represent the kinematics of the ionized and molecular gas. We do not show kinematic maps for the [P\,{\sc ii}], [S\,{\sc ix}] and \br\ lines (as done for flux maps) because these lines are detected only in a few regions and the \br\ kinematics is the same as that of \pb.

\begin{figure*}
 \centering
 \includegraphics[scale=0.64]{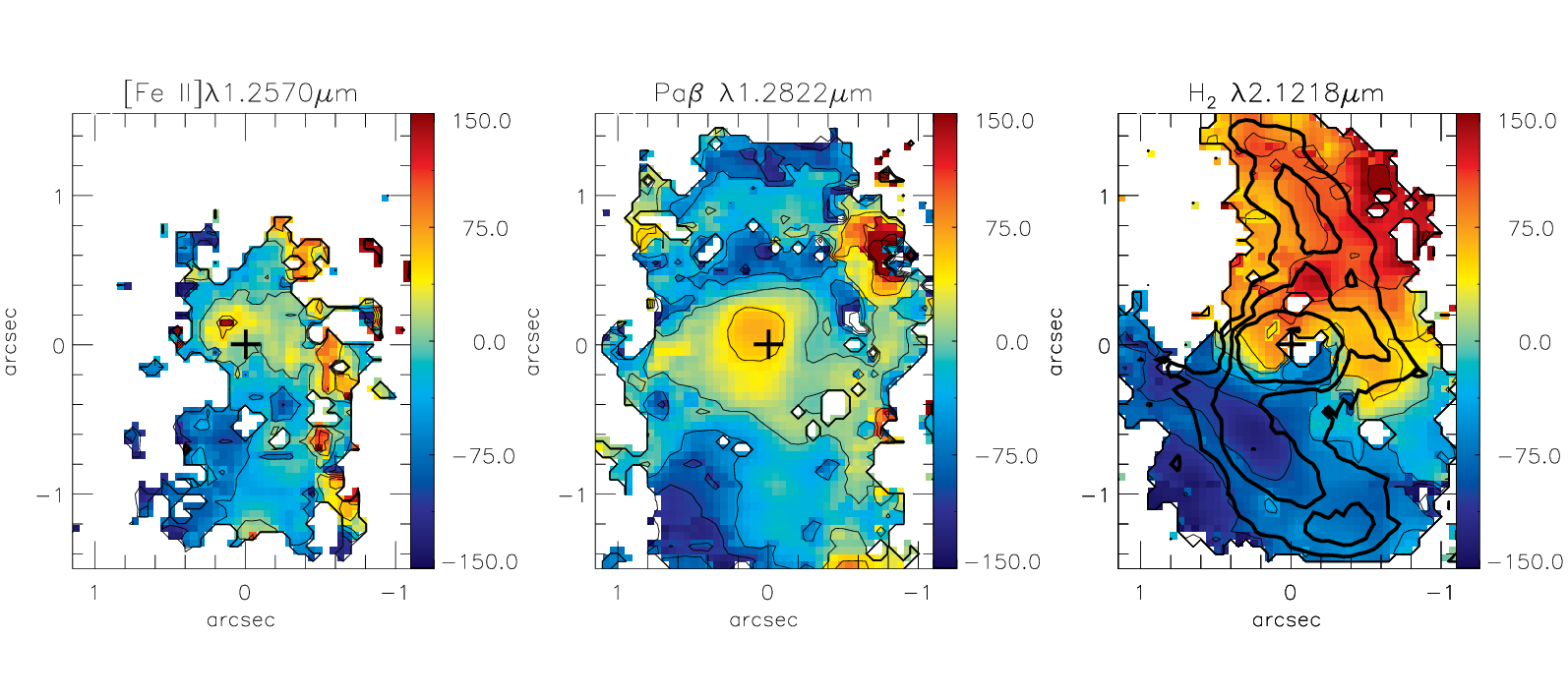} 
  \caption{Centroid velocity fields for the  \feii$\lambda1.2570\,\mu$m (left), \pb\ (middle) and H$_2\,\lambda$2.1218$\mu$m (left) emitting gas. The central crosses mark the position of the nucleus, North is up and East left and the thick contours overlaid to the \h2\ map are from the \h2\ flux image of Fig.\,\ref{flux}. Only the inner 1\farcs2 arcsec in the East-West direction is shown.} 
 \label{vel}  
 \end{figure*}

In Figure~\ref{vel} we present the velocity fields obtained from the centroid wavelength of [Fe\,{\sc ii}]$\lambda$1.2570$\mu$m (left panel), \pb\ (middle) and  H$_2\lambda$2.1218$\mu$m (right). The uncertainties in velocity are smaller than 15\,\kms\ for all emission lines at all locations of the field. White regions in this figure represent locations where the S/N was not high enough to allow the fitting of the line profiles.  We subtracted the heliocentric systemic velocity ($V_s\approx6636$\,\kms), obtained from the fitting of the \h2\ velocity field by a rotating disk 
model \citep[e.g.][]{mrk1066c,mrk1157}. The thick contours overlaid to the \h2\ velocity map are from the \h2$\lambda2.12$ flux image, shown in Fig.~\ref{flux}.  These maps show that the \feii\ and \pb\ emitting gas present similar velocity fields, with redshifts of about 40\,\kms\ at the nucleus and blueshifts of up to $-$150\,\kms\  both to the north and to the south. The \pb\ map also shows some redshifts in a region located at $\approx$1\farcs0 arcsec northwest of the nucleus, with velocities of 150\,\kms, similar to the values seen in the \h2\ velocity field at the same location.  
The H$_2$ velocity field shows an approximate rotation pattern, with the highest blueshifts, of up to $-$150\,\kms, to the southeast of the nucleus, and the highest redshifts, of up to 150\,\kms, to the northwest. The contours overlaid as thick lines are from the \h2\ flux distribution, showing that  deviations from the rotation pattern are associated with the spiral arms. In particular, a region of higher blueshifts than the surrounding  is seen inside the highest level contour delimiting the spiral structure to the south of the nucleus.

\begin{figure*}
 \centering
 \includegraphics[scale=0.64]{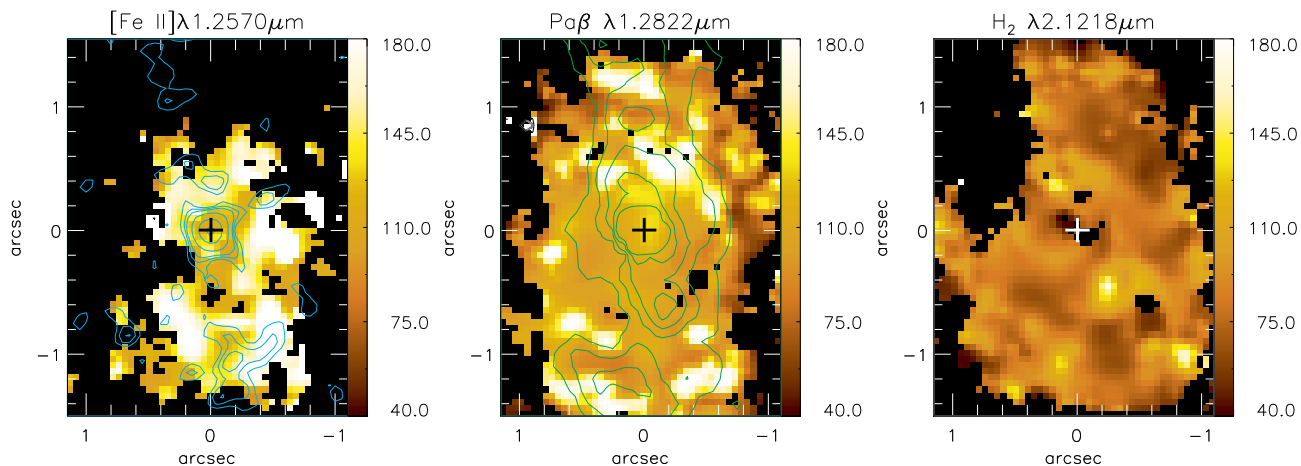} 
  \caption{$\sigma$ maps for the same emission lines of Fig.\,\ref{vel}.  
The cyan contours overlaid to the [Fe\,{\sc ii}] map are from the 3.6~cm radio-continuum image of \citet{schmitt01} and the green contours overlaid to the \pb\ map are from the [O\,{\sc iii}] image of \citet{schmitt03}.} 
 \label{sig}  
 \end{figure*}

In Figure~\ref{sig} we show the velocity dispersion maps for the [Fe\,{\sc ii}]$\lambda$1.2570$\mu$m (left panel), \pb\ (middle) and  H$_2\lambda$2.1218$\mu$m (right) emitting gas. Typical uncertainties in $\sigma$ are 15\,\kms. For the \feii\ and \pb\ the $\sigma$ values range from 40\,\kms\ to 180\,\kms, presenting the highest values in some knots located away from the nucleus. 
 In order to compare the \feii\ $\sigma$ and radio maps, we have overlaid the contours from the radio image on the \feii\ $\sigma$ map (cyan contours in Fig.\,\ref{sig}). This comparison shows that the highest $\sigma$ values are observed in regions surrounding the radio structures. 
The nuclear $\sigma$ for \feii\ and \pb\ are both of the order of $\approx$110\,\kms.
 The \h2\ $\sigma$ map is shown in the right panel of Fig.~\ref{sig} and presents smaller values than those observed for the \pb\ and \feii, with typical values being 75\,\kms. The smallest values ($\sim$60\,\kms) are observed in locations co-spatial with the spiral structure observed in the \h2\ flux map (Fig.~\ref{flux}), while the highest values of up to 130\,\kms\ are observed in some knots to south and southwest of the nucleus.

\subsection{Velocity Channel Maps}

\begin{figure*}
 \centering
 \includegraphics[scale=0.75]{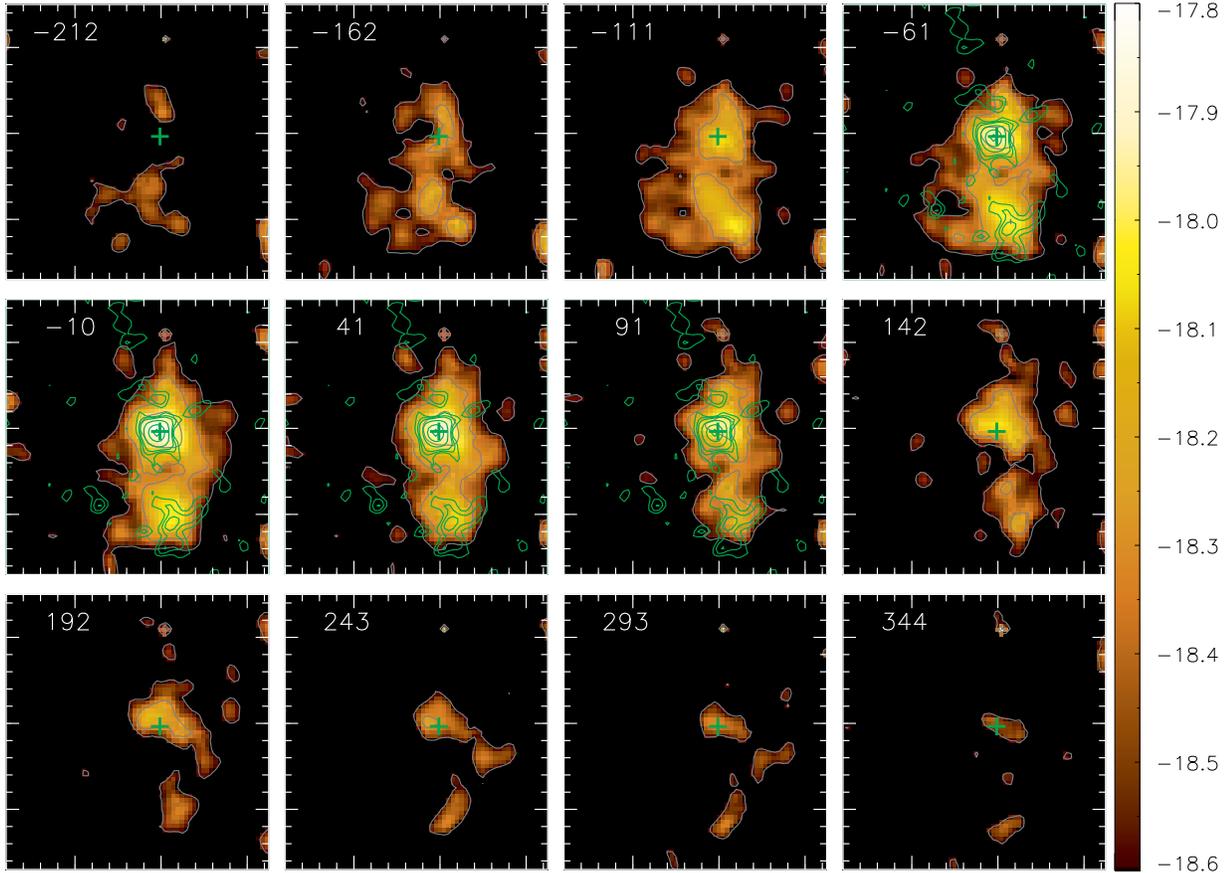} 
  \caption{Channel maps along the \feii\ emission-line profile for a velocity bin of 50\,\kms\ and centered at the velocity shown in the top-left corner of each panel. The green contours are from the 
radio image of \citet{schmitt03} and the cross marks the position of the nucleus.} 
 \label{slice_fe}  
 \end{figure*}

\begin{figure*}
 \centering
 \includegraphics[scale=0.75]{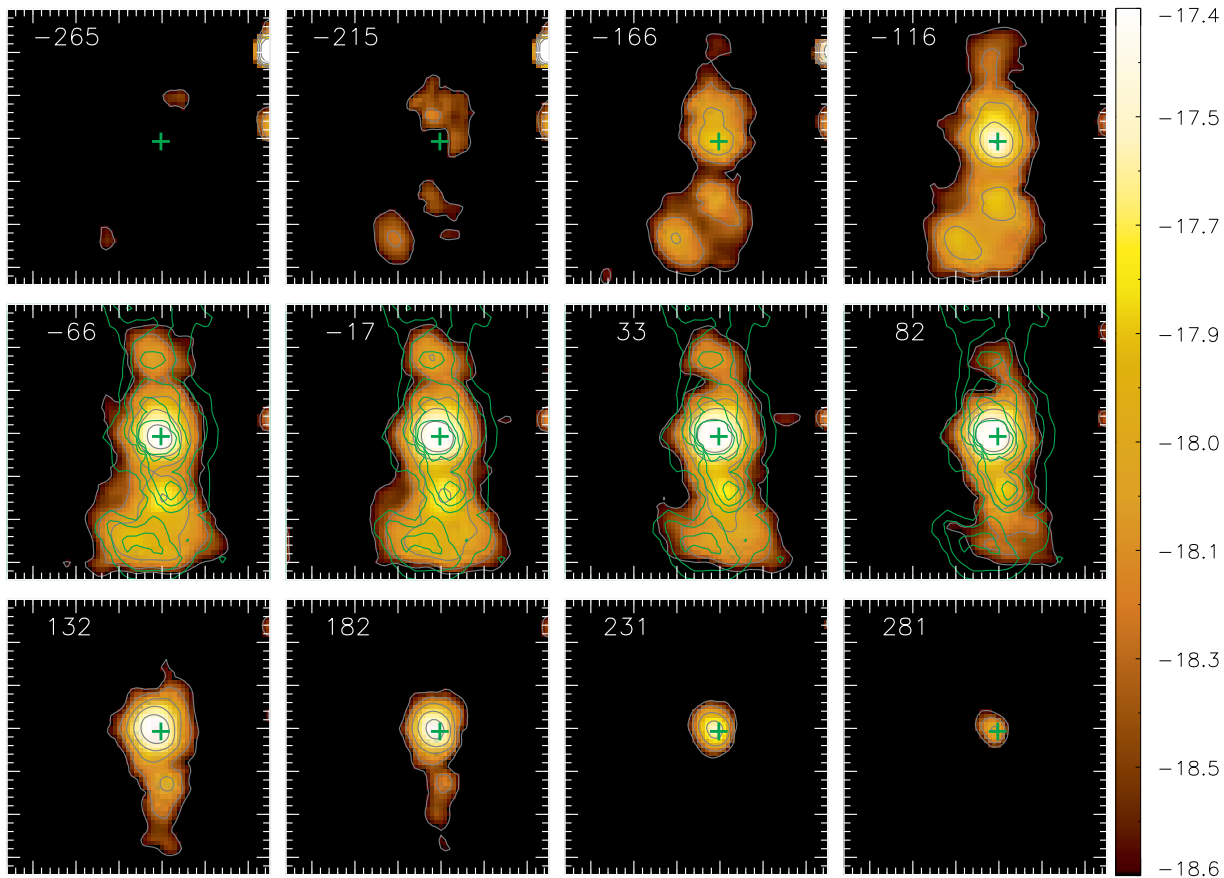} 
  \caption{Same as Fig.~\ref{slice_fe} for the  \pb\ emission line. The green contours are from the [O\,{\sc iii}] image from \citet{schmitt01}.} 
 \label{slice_pb}  
 \end{figure*}

\begin{figure*}
 \centering
 \includegraphics[scale=0.75]{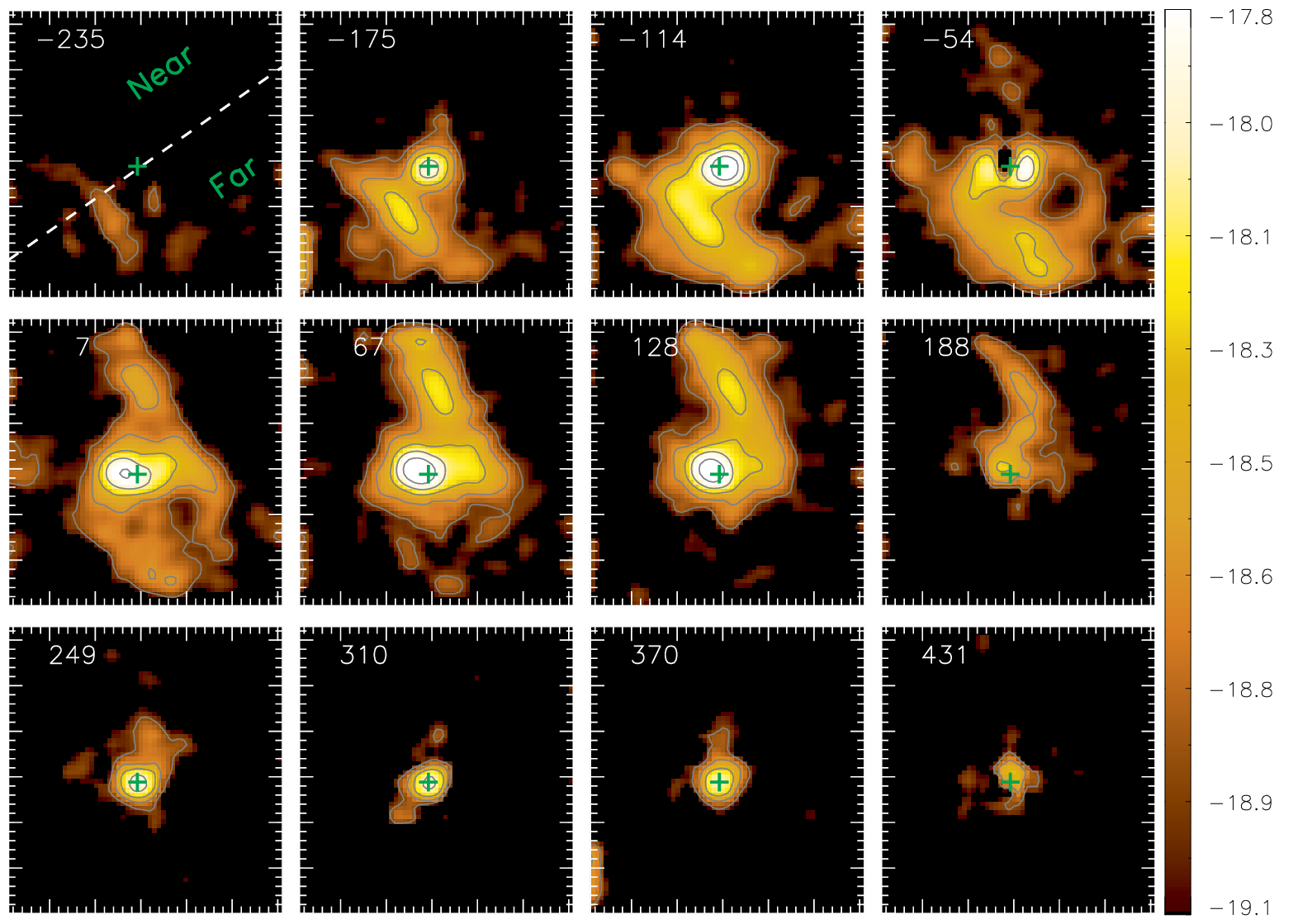} 
  \caption{Same as Fig.~\ref{slice_fe} for the  \h2\ emission line for a velocity bin of 60~\kms.} 
 \label{slice_h2}  
 \end{figure*}

In order to map the flux distributions at all velocities covered by the emission-line profiles, including the wings, we constructed channel maps along the profiles of the  \feii$\lambda1.2570\mu$m, \pb\ and \h2$\lambda2.1218\mu$m emission lines, shown in Figures~\ref{slice_fe}, \ref{slice_pb} and  \ref{slice_h2}, respectively. Each panel presents the flux distribution in logarithmic units integrated within the velocity bin centered at the velocity shown in the top-left corner (relative to the systemic velocity of the galaxy) and the central cross marks the position of the nucleus.

In Figure~\ref{slice_fe}, the channel maps along the \feii\ emission-line profile show the flux distributions integrated within velocity bins of 50~\kms (corresponding to two spectral pixels). 
The green contours overlaid to some panels are from the radio image of \citet{schmitt03}. The highest blueshifts of up to $-200$\,\kms\ and the highest redshifts, with similar velocities, are observed at the nucleus and from a region centered at $\sim$1\farcs0 to the south, where it is approximately coincident with the southern radio structure. At intermediate and zero velocities the \feii\ emission is observed from the nucleus down to $\approx$1\farcs4 to the south, still presenting the same two peaks, at the nucleus and  at 1\farcs0 to the south.

Figure~\ref{slice_pb} shows the channel maps for the \pb\ emitting gas for the same velocity bin as for \feii. The green contours overlaid to some panels are from the [O\,{\sc iii}] 
image of \citet{schmitt01}. At the highest blueshifts (velocities of $\sim-$200\,\kms) the \pb\ emission arises from three
structures: a ``blob'' at the nucleus, one  0\farcs7 to the south and another 1\farcs1 to the southeast (see channel maps centered at $-$166 and $-$215\,\kms) while the highest redshifts are observed at the nucleus and 0\farcs7 to the south. At velocities between $\approx-$160 and $\approx-$80\,\kms an additional structure is seen at 0\farcs8 north of the nucleus. At these velocities, the \pb\ emission is thus observed to both sides (north and south) of the nucleus, presenting 4 knots of enhanced emission, which are correlated to structures observed in the \oiii\ image, as can be seen from the \oiii\  contours overlaid to the panels at velocities ranging from $-$66 to 80\,\kms.

The channel maps along the \h2\ emission-line profile are shown in Figure~\ref{slice_h2} for a velocity bin of 60~\kms\ (corresponding to two spectral pixels). At the highest blueshifts ($-$235 to $-$175 \kms), the H$_2$ emission originates in a spiral arm extending for about 1\farcs0 and located southeast of the nucleus. For panels centered at velocities ranging from $-$175 to $-$54\,\kms, the spiral arm curves in the direction of the nucleus, resembling the arm seen in the \h2\ flux distribution of Fig.\,\ref{flux}. At zero velocities there is emission to both sides of the nucleus and as the velocities increase and reach positive values another spiral arm appears to the northwest. The highest redshifts of up to 400\,\kms\ are observed at the nucleus.

\section{Discussion}\label{discussion}

\subsection{Gaseous Excitation}\label{disc-excitation}

What is the origin of the near-IR lines of \feii\ and \h2\ from AGNs? This question has been investigated by several studies \citep[e.g.][]{black87,hollenbach89,forbes93,mouri94,maloney96,simpson96,larkin98,ardila04,ardila05,riffel06,mrk1066a,mrk1157,sb09,hicks09,sanchez09,ramos-almeida09}. In summary, the \h2\ 
emission lines can be excited by two mechanisms: (i) fluorescent excitation through absorption of soft-UV photons (912--1108 \AA) in
the Lyman and Werner bands \citep{black87} and (ii) collisional excitation due to the heating of the gas 
by shocks, due to interaction of a radio jet with the interstellar medium \citep{hollenbach89} 
or by X-rays from the AGN  \citep{maloney96}. The second mechanism, usually refereed as thermal processes, is also the responsible for the excitation of the near-IR lines of the \feii. Most of the studies above investigate the origin of both \h2\ and \feii\ using line ratios, such as [Fe\,{\sc ii}]$\lambda1.2570\mu$m/Pa$\beta$ and H$_2\,\lambda$2.1218\,$\mu$m/Br$\gamma$, and a common conclusion among these studies is that thermal processes dominate the \h2\ emission in the central region of active galaxies, while the fluorescent excitation can contribute only with a small fraction of the observed \h2\ emission. 

The main difficulty in the study of the excitation of the H$_2$ and \feii\ in AGNs regards the
distinction between X-ray and shock mechanisms. Recent detailed studies using integral field spectroscopy -- most of them by our AGNIFS group --  indicate that the \h2\ and \feii\ emitting gas have distinct flux distributions and kinematics, with the former being considered a tracer of the 
feeding of the AGN and the latter a tracer of its feedback \citep[e.g.][]{riffel06,mrk1066a,mrk1066c,mrk1157,sb09,sb10,hicks09,sanchez09}. 
This scenario is also favored in the case of Mrk\,79. The H$_2$ flux distribution in the shape of two spiral arms (Fig.\,\ref{flux}), observed also in the channel maps (Fig.\,\ref{slice_h2})  supports a location of the molecular gas in the plane of the galaxy. Additionally, the H$_2$ velocity field, shown in Fig.\,\ref{vel}, presents a clear rotation pattern with the southeast side approaching and the northwest side receding from us. The smaller velocity dispersion values observed for  H$_2$ (see Fig.\,\ref{sig}), relative to those of the ionized gas, also supports that the H$_2$ emitting gas is more restricted to the plane of the galaxy. The \feii\ emitting gas presents a more disturbed
velocity field, higher $\sigma$ values and a flux distribution well correlated with the radio structures,
indicating interaction with the radio jet. The presence of both blueshifts and redshifts to the south of the nucleus at the location of the southern radio structure, suggests that the  \feii\ emitting gas extends to high  galactic latitudes. The H\,{\sc i} recombination lines present a similar velocity field to that of \feii, but present smaller $\sigma$ values (intermediate values between those of \feii\ and H$_2$) and distinct flux distributions, being more associated to the \oiii\ emission than to the radio emission, as seen in the Figs.~\ref{flux} and \ref{slice_pb}. Thus, the \feii\ kinematics supports the presence of shocks contributing to its excitation.

The emission-line ratios can also be used to investigate the gas excitation. 
The \feii$\lambda1.25\mu$m/\pb\ and \h2$\lambda2.12\mu$m/\br\ emission-line ratios can be used to distinguish between Seyferts, LINERS and Starbursts \citep[e.g.][]{larkin98,ardila04,ardila05}. 
Seyfert nuclei have values in the range 0.6$\lesssim$\feii/\pb$\lesssim$2.0 and 0.6$\lesssim$\h2/\br$\lesssim$2.0, while Starburst galaxies have smaller values for both ratios and LINERs
have higher values \citep[e.g.][]{ardila05}. From the middle panel of
Fig.~\ref{ratio} it can be seen that the \feii/\pb\ ratio map presents typical Seyfert values for most of the
observed field. The only exception is the nucleus, where \feii/\pb$\approx0.15$, for which the \pb\ flux
can be increased due to contamination of the narrow component by the broad profile and thus, decreasing this ratio. The \h2/\br\ line ratio, shown in the right panel of Fig.~\ref{ratio}, presents most values in the range expected for Seyfert galaxies. Nevertheless, higher values are observed in the regions
where the spiral structures are seen in the H$_2$ flux distribution (Fig.~\ref{flux}) reaching \h2/\br$\approx4.5$.
These ratios can be understood by an enhancement of the \h2\  emission in the spiral arms due to the
increase of the gas density due to the spiral structure.

In a recent study,  \citet{dors12} built photoionization models considering a two-component
continuum, one to account for the Big Bump component peaking at 1 Ryd 
 and another
to represent the X-ray source that dominates the continuum emission at high energies in order to
reproduce the \feii$\lambda1.25\mu$m/\pb\ and \h2$\lambda2.12\mu$m/\br\  line ratios of AGNs. The authors compared
their models with the line ratios observed for a large sample of AGN from long-slit and IFU
spectroscopy. They concluded that typical Seyfert values for these ratios, as those observed for
Mrk\,79, are well reproduced by the model and concluded that the heating by X-rays produced by
active nuclei can be considered a common and very important mechanism of excitation of \feii\ and \h2\ lines. 
In the case of Mrk\,79, such conclusion  must be taken with caution since the \h2\ and \feii\ have distinct flux distribution and kinematics, suggesting that their emission originate from gas located at distinct regions  of the galaxy, as discussed above. This may also be the case of the extended emission of other Seyfert galaxy \citep[e.g.][]{riffel06,n4051,n7582,mrk1066a,mrk1066c,mrk1157,iau,sb09,sb10,sb10b}. 

 We have observed a good correlation between the \feii\ flux distribution and the radio emission, as well as an increase of the \feii\ $\sigma$ in locations co-spatial with the radio structures. This correlation is particularly clear in the channel maps shown in Fig.~\ref{slice_fe} for velocities ranging from $-61$ to $91$~km\,s$^{-1}$ -- enhancements in the flux  are observed at the locations of the two main radio structures: a hot spot at the nucleus and another at $\approx$\,1\farcs2 to the south of the nucleus.  This result suggest that the radio jet has an important role in the \feii\ emission, either by releasing iron from dust grains via shocks and increasing its abundance in  the gas phase as well as by enhancing the \feii\ emitting gas density due to compression produced by the radio jet.

The correlation between the NLR gas emission and radio structures has been questioned in detailed studies of nearby Seyfert galaxies. Examples are the studies of the NLR kinematics of NGC\,4151 \citep{kaiser00,das05} and NGC\,1068 \citep{das06}, using high-spatial resolution long slit spectra obtained with Space Telescope Imaging Spectrograph (STIS). The absence of any clear correlation between the optical emission lines and radio structures led the authors to conclude that there is no connection between the kinematics of the NLR and radio jets. The argument is that, even though these structures are approximately aligned as they originate in the same AGN, when the targets are close enough, high spatial resolution data
shows that there is no correlation between the  NLR gas kinematics and the radio jet. Although we believe that this may happen in some cases, we would like to point out that in the case of NGC\,4151, our study \citep{sb10} of the NLR kinematics using adaptive optics near-IR integral field spectroscopy, at similar spatial resolution to that of the above studies, have shown that there is interaction of the radio jet with ambient gas at low velocities. This finding was only possible because our 3D integral field spectroscopic data has a moderately high spectral resolution ($\sim$ 5000), allowing to inspect and separate the gas emission features at distinct velocities via channel maps. This separation allowed us to find  this interaction of the radio jet with gas at velocities close to systemic. At the high velocity channels, an outflowing component along the ionization cone dominates the gas emission, and does not show any correlation with the radio jet.

In the case of Mrk\,79, as pointed out above, even though it is more distant than NGC\,4151, the correlation with the radio jet is clearly seen in the channel maps as well as in the velocity dispersion maps.


\subsection{Mass of the ionized and molecular gas}\label{disc-mass}

The mass of the ionized and molecular gas can be estimated from the measured fluxes of the  Br$\gamma$ and H$_2\lambda2.1218\,\mu$m emission lines \citep{n4051,sb09,mrk1157}. Following  \citet{sb09} we  estimate the mass of ionized gas in the inner 
1.3\,$\times$\,1.3\,kpc$^2$ of Mrk\,79 by 
\begin{equation}
 \frac{M_{HII}}{\rm M_\odot}\approx3\times10^{17}\left(\frac{F_{\rm Br\gamma}}{\rm erg\,s^{-1}cm^{-2}}\right)\left(\frac{d}{\rm Mpc}\right)^2,
\end{equation}
where  $F_{\rm Br\gamma}$ is the integrated flux for the \br~emission line and $d$ is the distance to Mrk\,79. 
We have assumed an electron temperature $T=10^4$\,K and electron density $N_e=100\,{\rm cm^{-3}}$. 

The mass of hot molecular gas can be obtained as
\begin{equation}
 \frac{M_{H_2}}{\rm M_\odot}\approx5.0776\times10^{13}\left(\frac{F_{H_{2}\lambda2.1218}}{\rm erg\,s^{-1}\,cm^{-2}}\right)\left(\frac{d}{\rm Mpc}\right)^2,
\label{mh2}
\end{equation}
where  $F_{H_{2}\lambda2.1218}$ is the integrated flux for the \h2$\,\lambda2.1218\,\mu$m emission line
and it was assumed a vibrational temperature of T=2000\,K.

Integrating over the whole IFU field we obtain $F_{\rm Br\gamma}\approx2.7\times10^{-15}\,{\rm erg\,s^{-1}\,cm^{-2}}$ and
 $F_{H_{2}\lambda2.1218}\approx6.7\times10^{-15}\,{\rm erg\,s^{-1}\,cm^{-2}}$,  resulting in   $M_{HII}\approx7\times10^6\,{\rm M_\odot}$
  and $M_{H_2}\approx3\times10^3\,{\rm M_\odot}$.  The mass of molecular gas is 10$^3$ times smaller than of ionized gas but this 
H$_2$ mass represents only that of hot gas emitting in the near-IR \citep{sb09,mrk1157}. The values above are in good agreement with the ones found for other 
active galaxies \citep[e.g.][]{n4051,mrk1066a,sb09,mrk1157,mrk766}.

According to \citet{mazzalay12} the mass of the cold molecular gas can be estimated as  

\begin{equation}
\frac{M_{\rm H_2 cold}}{\rm M_\odot} \approx1174\left(\frac{L_{H_{2}\lambda2.1218}}{\rm L_\odot}\right),
\label{mh2}
\end{equation}
where $L_{H_{2}\lambda2.1218}$ is the luminosity of the H$_2$ line and results $ M_{\rm H_2 cold} \approx 2\times10^9 \,{\rm M_\odot}$ for Mrk\,79. The ratio between the cold and hot molecular mass is thus $7\times10^5$, being within the range of values found for active galaxies in general \citep{dale05,mazzalay12}.

\subsection{Kinematics of the emitting gas}\label{disc-gas}

The velocity field of the molecular gas is completely distinct from that of the ionized gas (Figure~\ref{vel}). Although both of them present 
blueshifts of up to $-150$\,\kms\ to the south of the nucleus, to the north the molecular gas presents redshifts of up to 150\,\kms\  in an approximate rotation pattern while the ionized gas presents instead similar blueshifts to those observed to the south. (The \feii\ does not present much extended emission to the north of the nucleus and thus it was not possible to study its centroid velocity at these locations). Adopting  the systemic velocity derived from the H$_2$ velocity field, we find redshifts of up to 50\,\kms\  around the nucleus for \feii\ and  \pb. 

A distinct kinematics for the \h2, \feii\ and \pb\ emitting gas is also supported by the $\sigma$ maps (Fig.~\ref{sig}), which show that, at most locations, the \h2\ presents the smallest $\sigma$ values ($\approx$\,70\,\kms), followed by \pb ($\approx$\,100--180\,\kms), with  \feii\ presenting the highest $\sigma$ values ($\approx$\,150--180\,\kms), which are correlated with the radio continuum emission. Differences in the velocity fields are also observed in the channel maps (Figs.~\ref{slice_fe}, \ref{slice_pb}  and \ref{slice_h2}). The \feii\ emission in all velocity channels is well correlated with the radio image, supporting an interaction between the radio jet and the emitting gas, which is probably being pushed away from the nucleus by the radio jet. 
 The accelerated particles which give origin to the radio emission also contribute to release the Iron (Fe) from dust grains, enhancing the \feii\ emission. In order to destroy the dust grains and release the Fe, fast shocks are necessary. The line-ratio \feii/\pii$\lambda 1.18\,\mu$m is useful to investigate the role of shocks for the \feii\ emission.  Ratios larger than 2 indicate
that shocks have passed through the gas destroying the dust grains, releasing the Fe and thus enhancing its observed abundance \citep{oliva01,sb09}. For Mrk\,79 the \feii/\pii\ is larger than this value for the extra-nuclear spectra, as seen in Table\,\ref{fluxes_j}, indicating the presence of shocks, while for the nucleus the ratio is $\approx$1.6, suggesting that much of the Fe is still locked  in dust grains.


The distinct kinematics for the  \h2, \pb\ and \feii\ emitting gas support that these gas phases are mostly located in distinct regions of the galaxy, 
These kinematics properties are similar to those we have found for other active  galaxies in previous studies \citep{ardila04,ardila05,riffel06,n4051,n7582,mrk1066c,mrk1157,sb99,sb09,sb10,sb10b}: the \h2\ kinematics is usually dominated by rotation in the galaxy disc while the \feii\ emitting gas has distinct kinematic components attributed to gas extending to high galactic latitudes, usually in outflow and in interaction  with a radio jet.  \citet[][e.g.]{hicks09} have studied the \h2\ kinematics from the inner $\approx$100~pc of a sample  of 9 Seyfert galaxies using the instrument SINFONI  at the ESO Very Large Telescope (VLT) and also concluded that it is dominated by rotation in a disc with typical radius of $\approx$30~pc and a comparable height.

\subsubsection{Feeding}\label{feeding}

The low $\sigma$ values, rotation pattern, the location in the plane of the galaxy, as well as the presence of spiral arms in the H$_2$ emitting gas supports the identification of H$_2$ as a tracer of the cold gas which feeds the active galactic nucleus.
The \h2\ velocity field  (right panel of Fig.~\ref{vel}) presents a rotation pattern with the northwest side receeding and the southeast side approaching. 
From this observed velocity field and assuming that the spiral arms of Mrk\,79 are trailing it can be concluded that  the near side of the galaxy is the northeast and the far side is the southwest as labeled in the top-left panel of  Fig.~\ref{slice_h2}. The rotation pattern and low $\sigma$ support a location of the molecular gas in the plane of the galaxy, and thus the spiral arm seen in blueshift to the southwest (far side) and the other seen in redshift to the northeast can be interpreted as tracing inflows towards the centre of Mrk\,79. Similar inflows along spiral arms at scales of tens to hundreds of parsecs have also been observed  in other galaxies in ionized gas at optical wavelengths \citep{fathi06,sb07,vandeVen09,muller11} and in molecular gas in the near-IR \citep{n4051,sanchez09,davies09,mrk1066c}. Following \citet{n4051}, we estimate the hot molecular gas mass inflow rate through a circular cross section as

\begin{equation}
\dot{M}_{H_2}=2m_p\,N_{H_2}\,v\,\pi r^2 n_{\rm arms},
\end{equation}
 where $v=v_{\rm obs}/{\rm sin} i$ is the velocity of the inflowing material in the plane of the galaxy, $v_{\rm obs}$ is the observed velocity, $i$ is the inclination of the disc, $r$ is the radius of the circular cross section, $N_{H_2}$ is the hot $H_2$ density, $m_p$ is the proton mass and $n_{\rm arms}=2$ is the number of spiral arms. 

$N_{H_2}$ can be estimated from the mass of hot molecular gas under the assumption that it is located in a disc with radius $r_d\approx1^{\prime\prime}=455\,pc$ (as observed in the H$_2$ flux map in Fig.~\ref{flux}) and height $h=30~pc$, as suggested by \citet{hicks09} for H$_2$ disks observed  in the inner few tens of pc of active galaxies by near-IR integral field spectroscopy. The resulting average hot H$_2$ density is 
$N_{H_2}=\frac{M_{\rm H_2}}{2m_p\pi r_d^2 h}=3.13\times10^{-3}\,{\rm cm^{-3}}$.    

The radius of the cross section can be obtained from Fig.~\ref{slice_h2} as the half of the width of the spiral arms $r\approx0\farcs3\approx135$~pc. Adopting $v_{\rm obs}\approx120$\,\kms  (Fig.~\ref{slice_h2}), we obtain $\dot{M}_{H_2}\approx2\times10^{-3}\, {\rm M_\odot\,yr^{-1}/sin} i$. Assuming that the inclination of the H$_2$ disk is $i\approx30^\circ$ \citep{paturel03}  the mass inflow rate is $\dot{M}_{H_2}\approx4\times10^{-3}\, {\rm M_\odot\,yr^{-1}}$. This value is within the range of molecular gas mass inflow rates found for other active galaxies \citep{fathi06,sb07,n4051,davies09,vandeVen09,sanchez09}.

\subsubsection{Feedback}\label{feedback}

As discussed above, the velocity fields for the ionized gas of Fig.~\ref{vel} are not compatible with pure rotation in the galactic disk as in the case of H$_2$. For \pb\, blueshifts are observed to both sides of the nucleus, although with less extended emission to the north. In the channel maps, to the south of the nucleus, the blueshifts observed for the ionized gas (both for \pb\ and \feii) are similar to those of the H$_2$ velocity field and thus we conclude that this emission is dominated by gas in rotation in the plane of the galaxy. On the other hand, there is also gas in redshift to the south, which cannot be in the plane and we interpret as due to an outflow projected against the far side of the galaxy. The emission in blueshift to the north, observed in the \pb\ channel maps may be interpreted as due to the counterpart of the southern outflow observed in redshift; the outflow to the north is thus oriented towards us.The redshifts to the north of the nucleus seen in the \pb\ velocity field have approximately the same velocities as those observed for the H$_2$ at the same location and is probably  due to emission from the disk.

We can estimate the mass outflow rate across a circular cross section with radius $r=$0\farcs7$\approx320$\,pc located at a distance of 1\farcs0 from the nucleus to the north, corresponding to a opening angle of $35^\circ$, obtained directly from the middle panel of Fig.~\ref{vel}, by
\begin{equation}
 \dot{M}_{\rm out} = \frac{m_p N_e v_{\rm obs} f A}{sin\,\theta},
\end{equation}
where $m_p$ is the proton mass, $N_e$ the electron density, $v_{\rm obs}$ is the observed velocity for the outflow, $f$ is the filling factor, $A$ is the area of the cross section and $\theta$ is the inclination of the cone axis relative to the plane of the sky \citep{mrk1066c}. Assuming $N_e=500\,{\rm cm^{-3}}$, $f=0.01$ and  $v_{\rm out}=75$~\kms, directly from  middle panel of Fig.~\ref{vel}, we obtain $\dot{M}_{\rm out}\approx3/{\rm sin\,\theta}~{\rm M_\odot\, yr^{-1}}$. \citet{bian02} pointed out that the inclination of the accretion disk of Mrk\,79 is $\approx$30.6$^\circ$ assuming that the Broad Line Region has a Keplerian velocity and using the FWHM from the broad component of H$\beta$, suggesting that the accretion disk has approximately the same orientation of the large scale disk \citep{paturel03}. Assuming that the cone axis is perpendicular to de accretion disk, we obtain $\theta\approx60^\circ$ and thus, the mass outflow rate would be $\dot{M}_{\rm out}\approx3.5~{\rm M_\odot\, yr^{-1}}$.  This value is in the range of mass outflow rates observed for other active galaxies  \citep{veilleux05,crenshaw07,barbosa09,sb10,n7582,mrk1066c}.

Following \citet{sb10}, we can use the above mass outflow rate to estimate the  kinetic power of the outflow by

\begin{equation}
\dot{E}\approx\frac{\dot {M}_{out}}{2}(v_{out}^2+\sigma^2),
\end{equation}
where $v_{out}=v_{obs}/sin\theta$ is the velocity of the outflowing gas
 and $\sigma$ is its velocity dispersion. Using $\sigma\approx150\,$km\,s$^{-1}$ (from Fig.\,\ref{sig})
and $v_{out}=v_{obs}/sin\theta=75{\rm km\,s^{-1}}/{\rm sin}\,60^\circ=75$\,km\,s$^{-1}$ we obtain 
$\dot{E} \approx 3.4\times10^{40}$ erg\,s$^{-1}$, which is in good agreement with those obtained for Seyfert galaxies and compact radio sources \citep{mrk1157,sb10,holt11,holt06,morganti05}.

\subsubsection{Mass accretion rate}

The accretion rate necessary to power the AGN at the nucleus of Mrk\,79 can be  obtained from

\begin{equation}
 \dot{m}=\frac{L_{\rm bol}}{c^2\eta},
\end{equation}
where $L_{\rm bol}$ is the nuclear bolometric luminosity, $\eta$ is the efficiency 
of conversion of the rest mass energy of the accreted material into radiation and $c$ 
is the light speed. 
The bolometric luminosity for Mrk\,79 is $L_{\rm bol}\approx2\times10^{44}\,{\rm erg\, s^{-1}}$ \citep{zhou10}. Assuming $\eta\approx0.1$ \citep[e.g.][]{frank02}, we obtain a mass accretion rate of $\dot{m}\approx3.5\times10^{-2}~{\rm M_\odot\, yr^{-1}}$.


We can compare the mass accretion rate with the mass inflow and outflow rates estimated in sections \ref{feeding} and \ref{feedback}, respectively.  The mass inflow rate for the hot molecular gas is one order of  magnitude smaller than the accretion rate.  As discussed in our previous studies, the total mass inflow rate may be larger than the one observed in hot molecular gas, since large amounts of cold molecular gas are observed in central region of active galaxies \citep[e.g.][]{mazzalay12,krips07,boone07,dale05}. On the other hand, the mass outflow rate in the NLR of Mrk\,79 is about two orders of magnitude larger than $\dot{m}$, a ratio comparable to those observed for other Seyfert galaxies, which indicates that most of the outflowing gas observed in the NLR of active galaxies does not originate in the AGN but in the interstellar medium surrounding the galaxy nucleus, which is pushed away by the AGN outflow \citep[e.g.][]{mrk1157}.

Finally, comparing the kinetic power with the bolometric luminosity we find that $\dot{E}$ is four orders of magnitude smaller than $L_{\rm bol}$, implying that only a small fraction of the mass accretion rate is transformed in kinetic power in the NLR outflows. This value is in reasonable agreement with the AGN feedback derived by \citet{dimatteo05} in simulations for the co-evolution of black holes and galaxies, in order to match the $M-\sigma$ relation and similar to those derived for Seyfert galaxies using optical integral field spectroscopy \citep{barbosa09}, after revising the filling factor and gas density values as in \citet{mrk1157}.

\section{Conclusions}\label{conclusions}

We have analyzed two-dimensional near-IR $J-$ and $K_l-$bands spectra from the inner $\approx$\,680\,pc radius of the Seyfert 1 galaxy Mrk\,79 obtained with the Gemini NIFS at a spatial resolution of $\approx$100~pc (0\farcs25) and velocity resolution of $\approx$\,35$-$45\,\kms. We have mapped the emission-line flux distributions and ratios, as well as the kinematics of the molecular and ionized emitting gas. The main conclusions of this work are:

\begin{itemize}

\item The coronal line emission is marginally resolved by our observations, being more extended to the north-south direction -- following the same orientation of the radio jet, reaching  $\approx$450\,pc from the nucleus. 

\item The \feii\ and \pii\ emission are well correlated with the radio emission, while the H\,{\sc i} recombination lines are better correlated with the [O\,{\sc iii}] emission.

\item We detected two nuclear spiral arms in the \h2\ flux distribution, extending by 680\,pc, one to the south of the nucleus and another to the north. These arms begin in a nuclear bar  with extent of$\approx$400\,pc oriented along the east-west direction.

\item The excitation of the \h2\ and \feii\ lines is due to heating of the gas by X-rays from the central AGN. A small contribution from shocks due to the interaction of the radio jet with the ISM may also contribute to the \feii\ emission, as evidenced by enhancements in the \feii\ flux and $\sigma$ at locations co-spatial with radio structures.

\item The kinematics of the \h2\  is dominated by rotation, with inflows along the nuclear spiral arms with a mass inflow rate of $\dot{M}_{H_2}\approx4\times10^{-3}\, {\rm M_\odot\,yr^{-1}}$, comparable to the mass accretion rate necessary to power the AGN of Mrk\,79 ( $\dot{m}\approx1.3\times10^{-3}~{\rm M_\odot\, yr^{-1}}$).

\item The kinematics of the ionized gas shows a rotation component to the south, which is probably in the galaxy plane, but also shows an outflow component, observed in redshift to the south and in blueshift to the north of the nucleus. The ionized gas mass outflow rate through a cross section with radius $\approx$\,320\,pc located at a distance of $\approx$455\,pc from the nucleus is $\dot{M}_{\rm out}\approx3.5~{\rm M_\odot\, yr^{-1}}$, indicating that most of this outflowing gas originates in the interstellar medium surrounding the galaxy nucleus, which is pushed by the AGN outflow.

\item The kinetic power of the outflow is four orders of magnitude smaller than the bolometric luminosity of the AGN of Mrk\,79, implying that only a small fraction of the mass accretion rate is transformed in kinetic power in the NLR outflows. 

\end{itemize}

\section*{Acknowledgements}
We thank an anonymous referee for useful suggestions which helped to improve the paper. This work is based on observations obtained at the Gemini Observatory, 
which is operated by the Association of Universities for Research in Astronomy, Inc., under a cooperative agreement with the 
NSF on behalf of the Gemini partnership: the National Science Foundation (United States), the Science and Technology 
Facilities Council (United Kingdom), the National Research Council (Canada), CONICYT (Chile), the Australian Research 
Council (Australia), Minist\'erio da Ci\^encia e Tecnologia (Brazil) and southeastCYT (Argentina).  
This research has made use of the NASA/IPAC Extragalactic Database (NED) which is operated by the Jet
 Propulsion Laboratory, California Institute of  Technology, under contract with the National Aeronautics and Space Administration.
This work has been partially supported by the Brazilian institution CNPq.

\label{lastpage}

\end{document}